# Fractal-like photonic lattices and localized states arising from singular and nonsingular flatbands

Yuqing Xie[1,6], Limin Song[1,6], Wenchao Yan[1], Shiqi Xia[1], Liqin Tang[1,2,4], Daohong Song[1,2,5], Jun-Won Rhim[3], Zhigang Chen[1,2]

[1] The MOE Key Laboratory of Weak-Light Nonlinear Photonics, TEDA Applied Physics Institute and School of Physics, Nankai University, Tianjin 300457, China
[2] Collaborative Innovation Center of Extreme Optics, Shanxi University, Taiyuan, Shanxi 030006, People's Republic of China
[3] Department of Physics, Ajou University, Suwon, 16499, Korea
[4] tanya@nankai.edu.cn, [5] songdaohong@nankai.edu.cn,
[6] These authors contributed equally to this work
Corresponding author: zgchen@nankai.edu.cn

**Abstract:** We realize fractal-like photonic lattices using cw-laser-writing technique, thereby observe distinct compact localized states (CLSs) associated with different flatbands in the same lattice setting. Such triangle-shaped lattices, akin to the first generation Sierpinski lattices, possess a band structure where singular non-degenerate and nonsingular degenerate flatbands coexist. By proper phase modulation of an input excitation beam, we demonstrate experimentally not only the simplest CLSs but also their superimposition into other complex mode structures. Furthermore, we show by numerical simulation a dynamical oscillation of the flatband states due to beating of the CLSs that have different eigenenergies. These results may provide inspiration for exploring fundamental phenomena arising from fractal structure, flatband singularity, and real-space topology.

## I. Introduction

The concept of flatband, as developed in condensed matter physics to describe certain geometrically frustrated lattices, was introduced into the realm of photonics by making an analogy between the electronic hopping and photonic coupling[1-4]. A flatband is a completely dispersion-free energy band covering the entire Brillouin Zone. Its localized nature is explained by the existence of special eigenmodes of the flatband - the so-called "compact localized states (CLSs)", which is usually stabilized by certain destructive quantum interference hosted by the given flatband lattice model[5]. Such CLSs have been realized in a number of experiments, including photonic Lieb lattices established by femtosecond laser-writing in glass[6, 7] and flatband Lieb and Kagome lattices optically induced in nonlinear crystals[8-10]. In recent years, flatband lattices have been employed to explore a variety of intriguing fundamental phenomena, such as those related to Landau-Zener Bloch oscillations[11], Moiré-lattice light localization[12], nontrivial topology and Floquet topological insulators[13-15], non-Hermiticity-induced PT-symmetry[16], and quantum distance and anomalous Landau levels[17]. Endeavors towards understanding the CLS completeness and Bloch wave singularities have unveiled the existence of unconventional flatband states – the noncontractible loop states (NLSs), as a manifestation of singular flatbands and nontrivial real-space topology[18-22].

Despite those efforts, most of research has focused on lattices with a flatband that is not isolated but rather "crossed" or "touched" by a dispersive band, such as Kagome and Lieb lattices. In addition, the previously established flatband lattices (either with a single or multiple flatbands) have a tight-binding Hamiltonian that exhibits only nondegenerate eigenvalue for each flatband. Recently, however, isolated flatbands[23] have attracted an increased attention[24, 25]. In particular, a fractal-like lattice structure was theoretically proposed, in which both isolated and non-isolated flatbands co-exist[26]. In fact, it was predicted that such a fractal-like lattice, as constructed from the well-known Sierpinski gasket fractal-network[26], could host multiple flatbands with even nonsingular but degenerate flatbands, and could lead to the understanding of nontrivial topological states in electronic fractal systems[15]. For a physically realizable photonic system, Floquet topological insulators in a fractal-dimensional lattice consisting of helical waveguides were also proposed[14]. Nevertheless, to our knowledge, a fractal or fractal-like lattice with nontrivial flatbands has not yet been realized in any experiment thus far, especially with respect to the characteristics of its flatband states.

In this Letter, we report experimental demonstration of fractal-like photonic lattices with coexisting three flatbands by use of a continuous-wave (cw) laser-writing technique to write waveguides one by one in a nonlinear crystal. Furthermore, under proper excitation condition, we observe compact localized flatband states distinct from each flatband. Specifically, one is singular nondegenerate and the other two are nonsingular degenerate, which is of particular interest since such a lattice structure, never been realized experimentally in any system before, can serve as a platform to explore the interplay between fractal structure, flatband singularity, and nontrivial topology. We measure both the intensity pattern and phase structure associated with the CLSs of each flatband, along with discussion about their superposition to form more complex localized structures or undergo dynamical oscillations.

## II. The lattice model and the band structure

In Fig. 1(a), different generations of a Sierpinski gasket for building the fractal-network are illustrated. In this work, we use the generation $l=1$ Sierpinski gasket as the unit cell to construct a fractal-like edge-centered triangular lattice (ECTL)[26]. The geometry of such a fractal-like ECTL is shown in Fig. 1(b), and a typical laser-written lattice structure corresponding to Fig. 1(b) is shown in

Fig. 1(c). Different from the common triangle lattice or the kagome lattice, there are six sites per unit cell. A light beam propagating in such a 2D lattice is governed by the following Schrödinger-type paraxial wave equation[8, 12, 27, 28]

$$i\frac{\partial \psi(x,y,z)}{\partial z} = -\left(\frac{1}{2k_0}\nabla_\perp^2 + \frac{k_0 \Delta n(x,y)}{n_0}\right)\psi(x,y,z) \equiv H_0 \psi(x,y,z), \quad (1)$$

where $\psi(x,y,z)$ is the electric field envelope of the probe beam, $z$ is the longitudinal propagation distance, $\nabla_\perp^2 = \partial_x^2 + \partial_y^2$ is the transverse Laplacian, $k_0$ is the wavenumber, and $n_0$ is the background refractive index of the medium. $\Delta n$ is the periodic refractive index change which is constructed by the cw-laser-writing technique in a bulk nonlinear photorefracive crystal

$$\Delta n(x,y) = -\frac{1}{2}n_0^3 \gamma_{33} \frac{E_0}{1+I(x,y)}, \quad (2)$$

where $I(x,y) = I_0 \left|\sum_{m,n} \exp\left[\left(-(x-x_m)^2 - (y-y_n)^2\right)/w^2\right]\right|^2$ is the intensity pattern of a series of superimposed Gaussian beams, which describes the fractal-like ECTL sturcutre. $\gamma_{33}$ is the electro-optic coefficient and $E_0$ is the bias field.

In Eq. (1), $H_0$ is a continuum Hamilton for light propagation in the photonic lattice. Under the tight-binding approximation which only considers the nearest-neighbor coupling of the waveguides, the corresponding discrete tight-binding Hamiltonian in the $k$-space can be written as [20, 26]

$$H_{TB} = t\begin{pmatrix} 0 & e^{ik_2} & e^{ik_3} & 1 & 1 & 0 \\ e^{-ik_2} & 0 & e^{-ik_1} & 1 & 0 & 1 \\ e^{-ik_3} & e^{ik_1} & 0 & 0 & 1 & 1 \\ 1 & 1 & 0 & 0 & 1 & 1 \\ 1 & 0 & 1 & 1 & 0 & 1 \\ 0 & 1 & 1 & 1 & 1 & 0 \end{pmatrix}, \quad (3)$$

where $t$ is the nearest-neighbor coupling constant, and we assume that $t=1$ for simplicity. $k_s = \mathbf{k} \cdot \mathbf{a}_s$ ($s=1,2,3$), $\mathbf{a}_1=(1,0)$, $\mathbf{a}_2=(1/2, \sqrt{3}/2)$, $\mathbf{a}_3=\mathbf{a}_2-\mathbf{a}_1$, $\mathbf{k}=(k_x,k_y)$. The key properties of the fractal-like ECTL are manifested in the tight-binding model $H_{TB}$. Solving the eigenvalue problem

$$H_{TB}|\beta_n,\mathbf{k}\rangle = \beta_n|\beta_n,\mathbf{k}\rangle, \quad (4)$$

in which the eigenvector is denoted as

$$|\beta_n,\mathbf{k}\rangle = (A_k, B_k, C_k, D_k, E_k, F_k)^T, \quad (5)$$

we obtain

$$\beta_n(\beta_n+2)^2(\beta_n^3 - 4\beta_n^2 + 6 - 2Q) = 0, \quad (6)$$

where $n$ ($n=1,\ldots,6$) is the band index, $Q=\sum_{s=1}^{3}\cos k_s$, and the components of $|\beta_n,\mathbf{k}\rangle$ represent the amplitudes of the eigenstate at A to F sites in the Bloch basis. The explicit form of eigenenergies of $H_{TB}$ (as a function of $k_x$ and $k_y$), which form the band structure (dispersion relation), are given by

$$\beta_1 = \frac{4}{3} + \frac{16}{3\Delta} + \frac{\Delta}{3}, \quad (7)$$

$$\beta_2 = \frac{4}{3} - \frac{8(1-\sqrt{3}i)}{3\Delta} - \frac{(1+\sqrt{3}i)\Delta}{6}, \quad (8)$$

$$\beta_3 = 0, \quad (9)$$

$$\beta_4 = \frac{4}{3} - \frac{8(1+\sqrt{3}i)}{3\Delta} - \frac{(1-\sqrt{3}i)\Delta}{6}, \quad (10)$$

$$\beta_5 = \beta_6 = -2, \tag{11}$$

where

$$\Delta = \left(-17 + 27Q + 3\sqrt{3}\sqrt{27Q^2 - 34Q - 141}\right)^{1/3}.$$

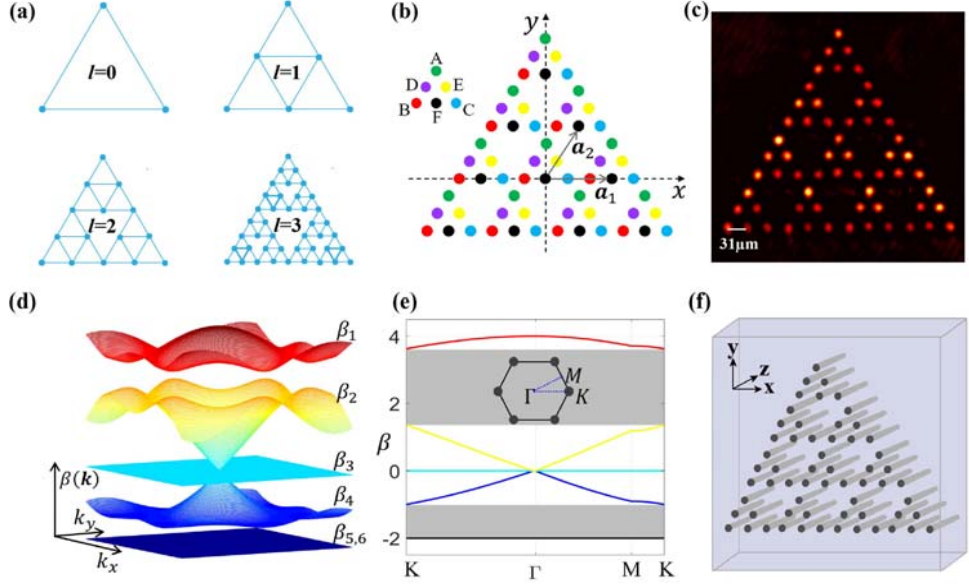

Fig. 1. (a) Illustration of different generation (indicated by $l$) Sierpinski gasket fractal-network. (b) Schematic of the fractal-like ECTL, where the building blocks (unite cells) are the first-generation ($l=1$) SPG fractal geometry. $a_1$ and $a_2$ are the primitive vectors in real space. (c) Experimental fractal-like photonic ECTL established with the cw-laser-writing technique. (d) Band structure from tight-binding model, where there are six bands which are marked by $\beta_{1\sim6}$. (e) Band structure plotted along high-symmetry points. The gray region denotes the band gaps. (f) 3D illustration of the fractal-like photonic ECTL, through which the excitation beam is launched along the longitudinal $z$-direction.

The calculated 2D band structure of the fractal-like ECTL in the first Brillouin zone (BZ) is displayed in Fig.1(d) and Fig.1(e). We find that the ECTL possesses three dispersive bands and three flat-bands, as was previously reported[26], including two degenerate nonsingular flatbands with the eigenvalues $\beta = -2$ and a nondegenerate flatband with $\beta = 0$. The two degenerate flatbands have the minimum propagation constant, so they are located in the bottom of the spectrum, completely isolated from other bands. We note that the ECTL is one of few realistic spinless models hosting degenerate flatbands in the sense that this model consists only of nearest hopping processes while other known degenerate flat band models, such as the third nearest hopping model on the Kagome lattice, usually contain unrealistic hopping processes[18, 20, 29]. On the other hand, the nondegenerate singular flatband touches two linearly dispersive bands at the center Γ point of the first BZ, forming a pseudospin-1 Dirac-like cone as that in Lieb and super-honeycomb lattices [19, 22, 30].

### III.    Results

Although flatband localization has been realized in a variety of lattices, such as Lieb[19], Kagome[21] and super-honeycomb lattices[22], all of which have only one non-degenerate singular flat band. In fact, there are very few studies on flatband photonic lattices with multiple flatbands or flatband degeneracy. Here, we use fractal-like photonic ECTL as an example to explore the properties of flatband states in lattices with multiple flatbands consisting of singular non-degenerate and nonsingular degenerate flatbands. According to the results obtained in Sec. II, the eigenstate corresponding to the non-degenerate flat-band $\beta_3$ can be obtained as

$$|\beta_3,\mathbf{k}\rangle = c_1 \left(1 - e^{ik_1}, e^{-ik_3} - 1, e^{ik_1} - e^{-ik_3}, e^{-ik_3} - e^{-ik_1}, 1 - e^{ik_3}, e^{ik_1} - 1\right)^T, \quad (12)$$

where $c_1 = \{4[3 - \cos k_x - 2\cos(k_x/2)\cos(\sqrt{3}k_y/2)]\}^{-1/2}$. One can choose $\alpha_\mathbf{k} = c_1^{-1}$, which makes $\alpha_\mathbf{k}|\beta_3,\mathbf{k}\rangle$ in the form of a finite sum of the Bloch phases. This leads to the amplitude of CLS-1 with the form

$$A_{0,\mathbf{R}}^{(\text{CLS-1})} = \left(\delta_{\mathbf{R},0} - \delta_{\mathbf{R},-a_1}, \delta_{\mathbf{R},a_3} - \delta_{\mathbf{R},0}, \delta_{\mathbf{R},-a_1} - \delta_{\mathbf{R},a_3}, \delta_{\mathbf{R},a_3} - \delta_{\mathbf{R},-a_1}, \delta_{\mathbf{R},0} - \delta_{\mathbf{R},a_3}, \delta_{\mathbf{R},-a_1} - \delta_{\mathbf{R},0}\right)^T. \quad (13)$$

From Eq. (13) we see that the CLS-1 occupies 12 lattice sites over three different unit cells, and it has the amplitudes +1 at A and E sites and −1 at B and F sites in the $\mathbf{R}=0$ unit cell, and +1 at C and F (B and D) sites and −1 at A and D (C and E) sites in the $\mathbf{R}=-a_1$ ($\mathbf{R}=a_3$) unit cell, as shown in Fig. 2(a1). The eigenvector $\alpha_\mathbf{k}|\beta_3,\mathbf{k}\rangle$ indicates that this nondegenerate flatband belongs to the singular flatband class because $\alpha_\mathbf{k}|\beta_3,\mathbf{k}\rangle = 0$ at $\mathbf{k} = 0$, and therefore the $N$ translated copies of CLS-1 cannot form a complete set, where $N$ is the number of unit cells in the system[18, 20, 29]. On the other hand, the eigenstates corresponding to the degenerate flat-band $\beta_{5,6}$ are given by

$$|\beta_5,\mathbf{k}\rangle = c_2\left(e^{ik_3} + e^{ik_2}, e^{ik_3} - 1, -1 - e^{ik_2}, -e^{ik_3} - e^{ik_2}, 0, 1 + e^{ik_2}\right)^T, \quad (14)$$

$$|\beta_6,\mathbf{k}\rangle = c_3\left(e^{ik_3} - e^{ik_2}, 1 + e^{ik_3}, -1 - e^{ik_2}, -1 - e^{ik_3}, 1 + e^{ik_2}, 0\right)^T, \quad (15)$$

where $c_2 = [2(5 + 2\cos k_1 + 2\cos k_2 - \cos k_3)]^{-1/2}$, $c_3 = [2(5 + 2\cos k_2 + 2\cos k_3 - \cos k_1)]^{-1/2}$ are normalization coefficients. Equations (14) and (15) indicate the flatband modes have zero amplitude on the E and F sublattices. Then, we can obtain the amplitudes of the CLS-2 and CLS-3

$$A_{0,\mathbf{R}}^{(\text{CLS-2})} = \left(\delta_{\mathbf{R},-a_3} + \delta_{\mathbf{R},-a_2}, \delta_{\mathbf{R},-a_3} - \delta_{\mathbf{R},0}, -\delta_{\mathbf{R},0} - \delta_{\mathbf{R},-a_2}, -\delta_{\mathbf{R},-a_3} - \delta_{\mathbf{R},-a_2}, 0, \delta_{\mathbf{R},0} + \delta_{\mathbf{R},-a_2}\right)^T, \quad (16)$$

$$A_{0,\mathbf{R}}^{(\text{CLS-3})} = \left(\delta_{\mathbf{R},-a_3} - \delta_{\mathbf{R},-a_2}, \delta_{\mathbf{R},0} + \delta_{\mathbf{R},-a_3}, -\delta_{\mathbf{R},0} - \delta_{\mathbf{R},-a_2}, -\delta_{\mathbf{R},0} - \delta_{\mathbf{R},-a_3}, \delta_{\mathbf{R},0} + \delta_{\mathbf{R},-a_2}, 0\right)^T, \quad (17)$$

as shown in Figs. 2(b1) and 2(c1). One can note that $c_2^{-1}|\beta_5,\mathbf{k}\rangle$ and $c_3^{-1}|\beta_6,\mathbf{k}\rangle$ become zero at $\mathbf{k} = (\pi, \pi/\sqrt{3})$ and $\mathbf{k} = (0, 2\pi/\sqrt{3})$ respectively. Due to this property, one might conclude that these two degenerate flatbands are singular ones. However, due to the complete degeneracy between these two bands, one can make a nonsingular set of eigenvectors for those degenerate flatbands from the linear combinations between $c_2^{-1}|\beta_5,\mathbf{k}\rangle$ and $c_3^{-1}|\beta_6,\mathbf{k}\rangle$ because the singular momenta of $c_2^{-1}|\beta_5,\mathbf{k}\rangle$ and $c_3^{-1}|\beta_6,\mathbf{k}\rangle$ are distinguished from each other. Therefore, these two degenerate flatbands belong to the nonsingular flatband category.

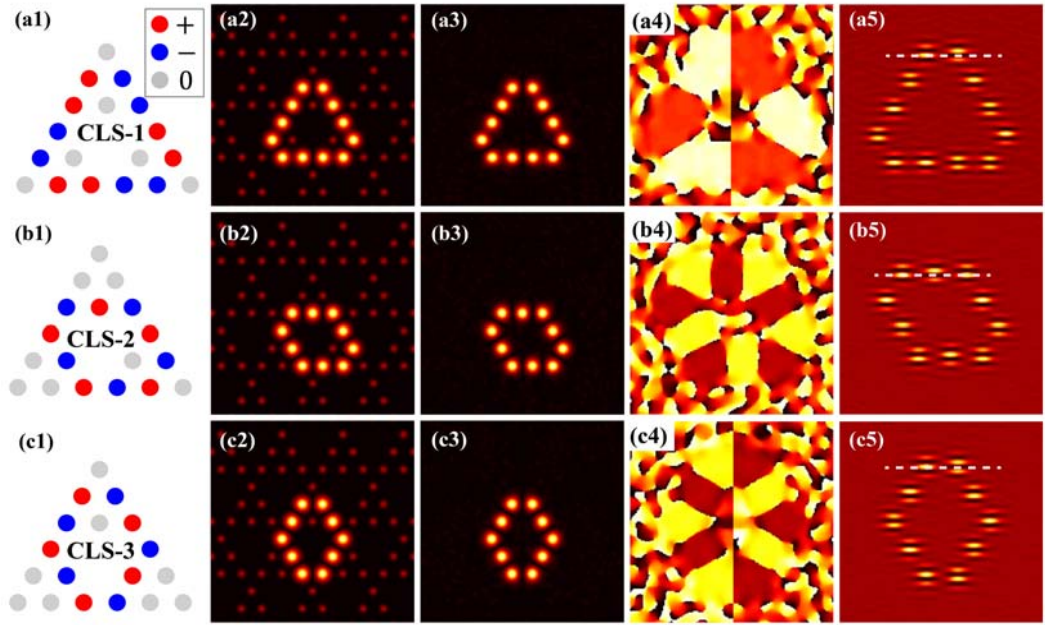

Fig. 2. Numerical results showing three different fundamental CLSs in fractal-like photonic lattices. (a1) Illustration of the CLS-1 for the nondegenerate flatband $\beta=0$, where zero amplitudes sites are denoted by gray color, and red and blue colors represent non-zero equal amplitudes with opposite phase. (a2) Superimposed pattern of the CLS-1 with the fractal-like lattice at input. (a3) Output transverse intensity pattern of the CLS-1 after propagating 1 cm through the lattice. (a4) and (a5) are the zoomed phase distribution and interferogram at output, confirming the phase structure shown in (a1). The layouts of (b1)-(b5) and (c1)-(c5) are the same as in (a1)-(a5), except that they are for the CLS-2 and CLS-3 from the two degenerate flatbands with $\beta=-2$.

First, we present numerical simulation and experimental observation of the fundamental CLSs propagating through the fractal-like photonic ECTL. When the CLS profiles shown in Figs. 2(a1)–2(c1) are used as a probe beam at the input to the lattice, they remain localized and intact during propagation through the lattice due to excitation of the corresponding fundamental flatband eigenmodes, as shown in Figs. 2(a3)–2(c3). The corresponding phase structures of the output also match that of CLSs, as can be seen from the phase structure of the output shown in Figs. 2(a4)–2(c4), as well as the corresponding interferograms [Figs. 2(a5)–2(c5)] of the output beam obtained by interfering the output beam with an inclined plane wave. According to the interference stripes marked by the white dashed lines, one finds that the initial out-of-phase structure is well preserved. These results clearly indicate that the CLSs obtained from the tight-binding model are indeed the flatband eigenstates associated with the three flatbands.

To observe the above distinct CLSs in experiments, we use the cw-laser-writing technique to construct the finite-sized fractal-like photonic ECTL in a nonlinear crystal, as established previously already to observe the flatband states in Lieb lattice[19], super-honeycomb lattice[22] and Kagome lattice[21]. For the writing process, an ordinarily polarized beam from a solid-state laser (532 nm) is sent through the biased SBN:61 photorefractive crystal after passing through a spatial light modulator (SLM) and a 4F system. The beam is configured to exhibit quasi-nondiffracting in the 10-mm-long crystal which writes the waveguides with controllable position. An experimentally established fractal-like photonic ECTL with a lattice spacing of 31 μm is shown in Fig. 1(c), which remains invariant during the entire writing and probing process due to the photorefractive "memory effect"[19]. For probing, an extraordinarily polarized beam with appropriate initial phase modulation to match that of the CLSs is launched into the lattice to observe localized propagation of different CLSs. Indeed, when the probe beam with the CLS-1 pattern (Fig. 3(a1)) is sent into the lattice, the probe beam remains localized in the initially excited waveguides (Fig. 3(a2)) after 10-mm propagation, and its phase structure can be seen from the interferograms obtained by interfering the output with a broad beam (quasi-plane wave) as shown in Fig. 3(a3). As an example, we show that the bottom two sites are in phase with each other but their phase is opposite with the neighboring site in the dotted white box, characterizing the phase structures of CLS-1. Likewise, when the probe beam is shaped into the pattern of CLS-2 (Fig. 3(b)) and CLS-3(Fig. 3(c)), both output patterns through the lattice remain localized, and the corresponding interferograms indicate that their phase structure is also preserved [Fig.3(b1-b3) and Fig.3(c1-c3)]. These experimental results agree well with the corresponding simulation results displayed in Fig.2. Furthermore, for a direct comparison, we show that if the spots in the probe beam are all in phase (Fig. 3(d1)) which does not match the phase structure of the eigenmode CLS-1, the input beam simply cannot be localized, but rather it couples to other neighboring waveguides as shown in Fig. 3(d2) and Fig. 3(d3).

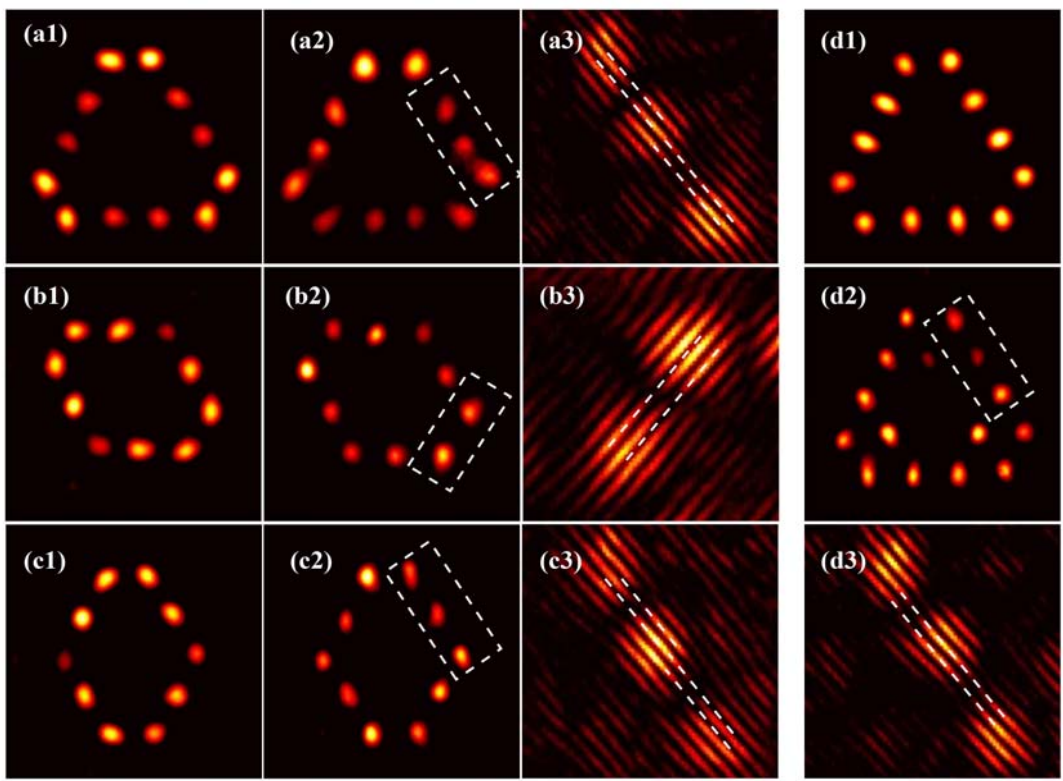

Fig. 3. Experimental demonstration of three different fundamental CLSs in fractal-like photonic lattices corresponding to those in Fig.2. Shown in (a1-a3) are the probe beam for excitation of CLS-1 at input, its output through the lattice, and its zoom-in interferogram to show the phase structure in the region highlighted by the white dotted box in (a2). Shown in (b1-b3) and (c1-c3) are the corresponding results for observation of CLS-2 and CLS-3, respectively. (d1-d3) show a typical example (corresponding to CLS-1) when the excitation beam has in-phase relation for all sites, so it cannot remain localized during propagation (d2).

Next, we study the propagation dynamics of linear superposition of the CLSs from the degenerate nonsingular flatbands. For the degenerate flat-bands with $\beta = -2$, we can find another form of the flatband eigenstates by appropriately combining the fundamental CLSs from $|\beta_5,k\rangle$ and $|\beta_6,k\rangle$,

$$|\beta',k\rangle = c_4\left(e^{ik_2} + e^{ik_3}, -1 - e^{ik_3}, e^{ik_2} - 1, 0, -e^{ik_2} - e^{ik_3}, 1 + e^{ik_3}\right)^T, \quad (18)$$

where $c_4 = \left[2(5 + 2\cos k_1 + 2\cos k_3 - \cos k_2)\right]^{-1/2}$, which can be obtained directly from the symmetry of the ECTL lattice, as displayed in Fig. 4(b3) and labeled as CLS-4. The amplitude and phase distributions of these superimposed CLSs are the same and their directions form an angle of $120°$ with each other. It is noted that these CLS superpositions are not all linearly independent, as one of the three can be represented by a linear combination of the other two, as depicted in Fig. 4(a1)-4(a3). Note that one cannot obtain a single rhombic CLS-4 from the rhombic CLS-2 and CLS-3. In other words, it is impossible to obtain a single rhombic CLS by the linear combination of other kinds of rhombic CLSs. The linear combination of these three CLSs leads to a super-CLS (Fig. 4(b4)) with the same shape but different phase distribution as compared with the CLS-1 from the singular flatband $\beta = 0$. Obviously, the super-CLS with the alternating out-of-phase distribution is also an eigenmode of the nonsingular flatband $\beta = -2$. In our experiment, these localized superposition mode 2 CLS-4s and super-CLS are also observed in the ECTL, as shown in (Fig. 4(a4)) and (Fig. 4(b5)). In contrast, when the probe beam is changed into in-phase structure, the output pattern cannot be localized anymore with energy coupling to other nearest sites (see Fig. 4(a5) and Fig. 4(b6)). These results show clearly that the superposition of fundamental CLSs associated with the degenerate flatbands can lead to other complex localized structures.

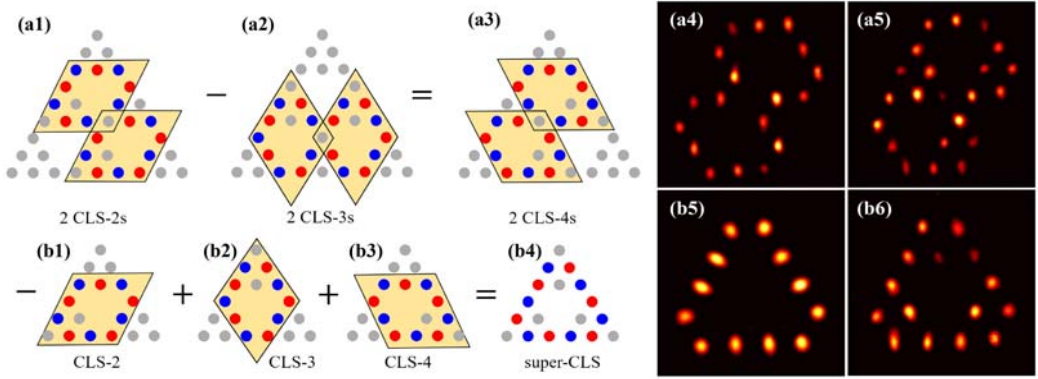

Fig. 4. Experimental observation of flatband superposition states associated with two degenerate flatbands. (a1)-(a3) Illustration of the rhombic CLS-4s in (a3) which is obtained by a linear combination of CLS-2s in (a1) and CLS-3s in (a2). (a4)-(a5) shows typical experimental results of the observed output intensity patterns corresponding to (a3), when the excitation beam is modulated with out-of-phase relation for adjacent sites (a4) or simply all in-phase relation for direct comparison. (b1)-(b6) Corresponding results for the super-CLS obtained by linear superposition of three fundamental CLSs, which displays same intensity but different phase distribution as compared with the CLS-1 in Fig. 2(a1) that belongs to the singular nondegenerate flatband.

## IV. Dynamical oscillation of flatband superposition states

Finally, we explore the dynamics of the flatband states by superposition of the CLSs from both the singular nondegenerate flatband ($\beta=0$) and the nonsingular degenerate flatbands ($\beta=-2$). Specifically, the superposition flatband state is constructed by a linear combination of CLS-1 and the super-CLS as shown in Fig.5(a), and the simulation results for long propagation distance to z=340 mm are summarized in Figs. 5(b1)-5(b6). We find that during propagation the initial excitation leads to intensity distribution always in a finite number of sites, although there is oscillation between these sites that results in shifting of the intensity peak: The minimum is the number of sites occupied by the superposition state after diffraction cancellation, whereas the maximum is the number of sites occupied by the union set of the two independent CLSs before diffraction cancellation. This oscillation can be seen more clearly from the side-view propagation of the superimposed state across one period as shown in Fig. 5(c). If we divide a period of propagation distance (around 340 mm for the parameters used) into two parts, then the evolution of the second part is a z-reversal of the first part along the propagation direction. The periodic oscillation of the superimposed mode is due to the mode beating between two different flatband CLSs which have different propagation constants. The difference between the propagation constants of the two CLSs determines the oscillation period. Such long distance dynamical oscillation of the CLS superposition states, however, cannot be realized in our current experimental condition due to the limited length of the photorefractive crystal.

Before closing, we would like to point out that, this current work has focused only on demonstration of the fundamental CLSs and their superpositions in fractal-like photonic lattices. However, there are many open issues remain to be explored. For example, we already showed that one can encounter with flat bands with a "plane degeneracy", where a flat band touches another flat band all over the Brillouin zone. How the plane degeneracy differs from the point degeneracy? Can the flat-band degeneracy give rise to NLSs, also the robust boundary modes (RBMs) [15]? Based on the tailored edge-centered triangle lattice constructed from the Sierpinski gasket, several typically direct and indirect manifestations of the NLSs, such as the (straight and wavy) line states, the RBMs, and the flat-band loop states in Corbino-shaped geometry can be realized. We find that the necessity of the existence of the NLSs depends on the selection of the spanning sets of the degenerate flat band, as well as the singularity of the flat band. These results on NLS formation along with their implication to real-space topology will be reported elsewhere.

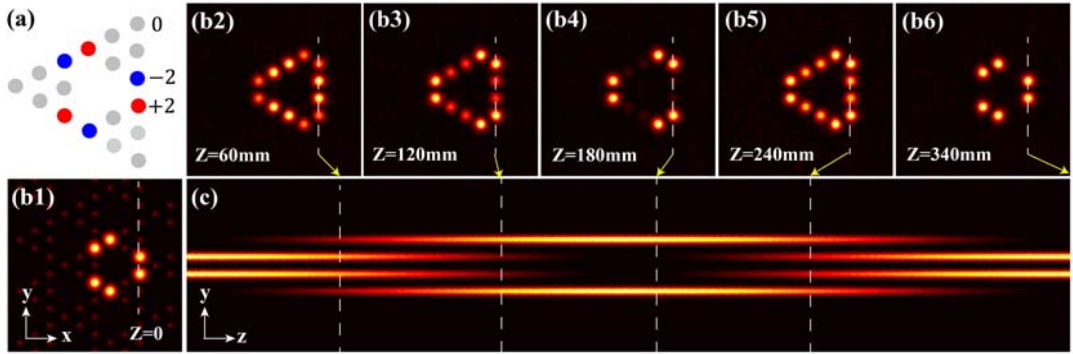

Fig. 5. Numerical results of dynamical oscillation of the localized states from superposition of CLSs belonging to different flatbands. (a) The flatband superposition state is depicted as colored sites obtained from the linear superposition of CLS1 with $\beta=0$ and super-CLS with $\beta=-2$. (b) Snapshots of output transverse intensity patterns at different propagation distances. (c) "Side view" evolution of the mode profiles during dynamical oscillation taken along the dashed line in (b).

## V. Conclusion

In conclusion, we have experimentally established the photonic fractal-like ECTL lattices with both singular nondegenerate and nonsingular degenerate flat-bands by the cw-laser-writing technique, and observed different types of CLSs and their superposition states. In addition, the dynamical oscillation of superposition states arising from flatbands with different eigenvalues is also numerically verified in such lattices. The fractal-like photonic lattices presented here may provide inspiration for exploring the fundamentals and applications of flat-band physics and fractal photonic topological insulators in other systems beyond photonics. In particular, our results may bring about new opportunities to explore many intriguing phenomena such as the NLSs and real-space topology, nonlinear flatband states and fractal flatband topology in such unconventional lattices.

**Acknowledgement:**

This work was supported by National Key R&D Program of China (2017YFA0303800); National Natural Science Foundation of China (11922408, 11674180 and 91750204); the Fundamental Research Funds for the Central Universities(63213041), PCSIRT (IRT_13R29), 111 Project (No. B07013) in China. J.W.R. was supported by the National Research Foundation of Korea (NRF) Grant funded by the Korea government (MSIT) (Grant No. 2021R1A2C1010572) and (Grant No. 2021R1A5A1032996).